\documentclass[12pt]{article}
\usepackage{graphicx}

\def\virginia{Department of Physics, University of Virginia, Charlottesville, VA 22904, USA\\Laboratori Nazionali di Frascati, INFN, Frascati, Italy}
\def\tufts{Department of Physics and Astronomy, Tufts University, Medford, MA 02155 USA}
\def\supportva{\footnote{Work supported by U.S. D.O.E. grant DE-FG02-01ER4120}}

\def\Title#1{\begin{center} {\Large #1 } \end{center}}
\def\Author#1{\begin{center}{ \sc #1} \end{center}}
\def\Address#1{\begin{center}{ \it #1} \end{center}}

\newenvironment{Abstract}{\begin{quotation}  }{\end{quotation}}
\newenvironment{Presented}{\begin{quotation} \begin{center} 
             PRESENTED AT\end{center}\bigskip 
      \begin{center}\begin{large}}{\end{large}\end{center} \end{quotation}}

\def\beq{\begin{equation}}
\def\eeq#1{\label{#1}\end{equation}}
\def\eeqn{\end{equation}}

\def\beqa{\begin{eqnarray}}
\def\eeqa#1{\label{#1}\end{eqnarray}}
\def\eeqan{\end{eqnarray}}

\let\bar=\overbar

\def\Dslash{\not{\hbox{\kern-4pt $D$}}}
\def\dslash{\not{\hbox{\kern-2pt $\del$}}}

\def\msb{{\bar{\ssstyle M \kern -1pt S}}}

\begin{document}
\begin{titlepage}

\vfill
\Title{Probing Spin Dependent Parton Distributions Through Deeply Virtual Processes}
\vfill
\Author{Gary R. Goldstein} 
\Address{\tufts}
\Author{Simonetta Liuti\supportva}
\Address{\virginia}
\vfill
\begin{Abstract}
Spin and transverse momentum dependent parton distributions - Generalized Parton Distributions (GPDs)
- are at the interface between the QCD structure of the hadrons and observable quantities. The GPDs are linear superpositions within helicity amplitudes.
 The amplitudes are probed in high energy leptoproduction processes through angular dependent cross sections and polarization asymmetries. Phenomenological extraction of the amplitudes and the distributions is a challenging task. We present observables that connect particularly with the chiral odd quark-nucleon helicity amplitudes for Deeply Virtual $\pi^0$ Production.
\end{Abstract}
\vfill
\begin{Presented}
CIPANP2015, May 19-24,  2015 (Vail CO) 
\end{Presented}
\vfill
\end{titlepage}
\def\thefootnote{\fnsymbol{footnote}}
\setcounter{footnote}{0}
%



The spin of the nucleon and all the hadrons, depends on the distributions of spin and orbital angular momenta (OAM) of the fundamental constituents, quarks and gluons. Since the discovery that much of the {\it helicity} of the proton is not associated with the quarks' helicity (through the measurement of the parton distribution function (pdf) $g_1^q (x)$), a very rich field of phenomenology has developed to account for the angular momentum. The quark and gluon field correlations in the nucleon are indirectly measurable through electroproduction processes. To investigate the angular momenta associated with the quark and gluon fields within QCD, the transverse momentum distributions (TMDs) and the Generalized Parton Distributions (GPDs) (dependent on momentum transfer) were developed. These go beyond the pdf's  measured in deep inelastic scattering and provide a window into a 3-dimensional picture of the angular momentum structure.

Of special interest among pdf's are the nucleon's {\it transversity} structure functions, $h_1(x)$, for the probability of finding a definite transversity quark inside a transversely polarized nucleon. These have been difficult to extract from experiment. Being chirally odd, they can be observed in either Semi Inclusive Deep Inelastic Scattering (SIDIS), where they are convoluted with fragmentation functions, or in the Drell-Yan process in conjunction with another chiral-odd partner. They also contribute to exclusive electroproduction processes, particularly Deeply Virtual Meson Production (DVMP), through chiral odd GPDs. In particular, the {\it transversity} GPD $H_T^q(x,\xi,t)$ has the limiting form $H_T^q(x,0,0)=h_1^q(x)$. Hence, from the DVMP processes, transversity can be determined. This will be a main focus of this presentation. 

\begin{figure}
\includegraphics[width=6.5cm]{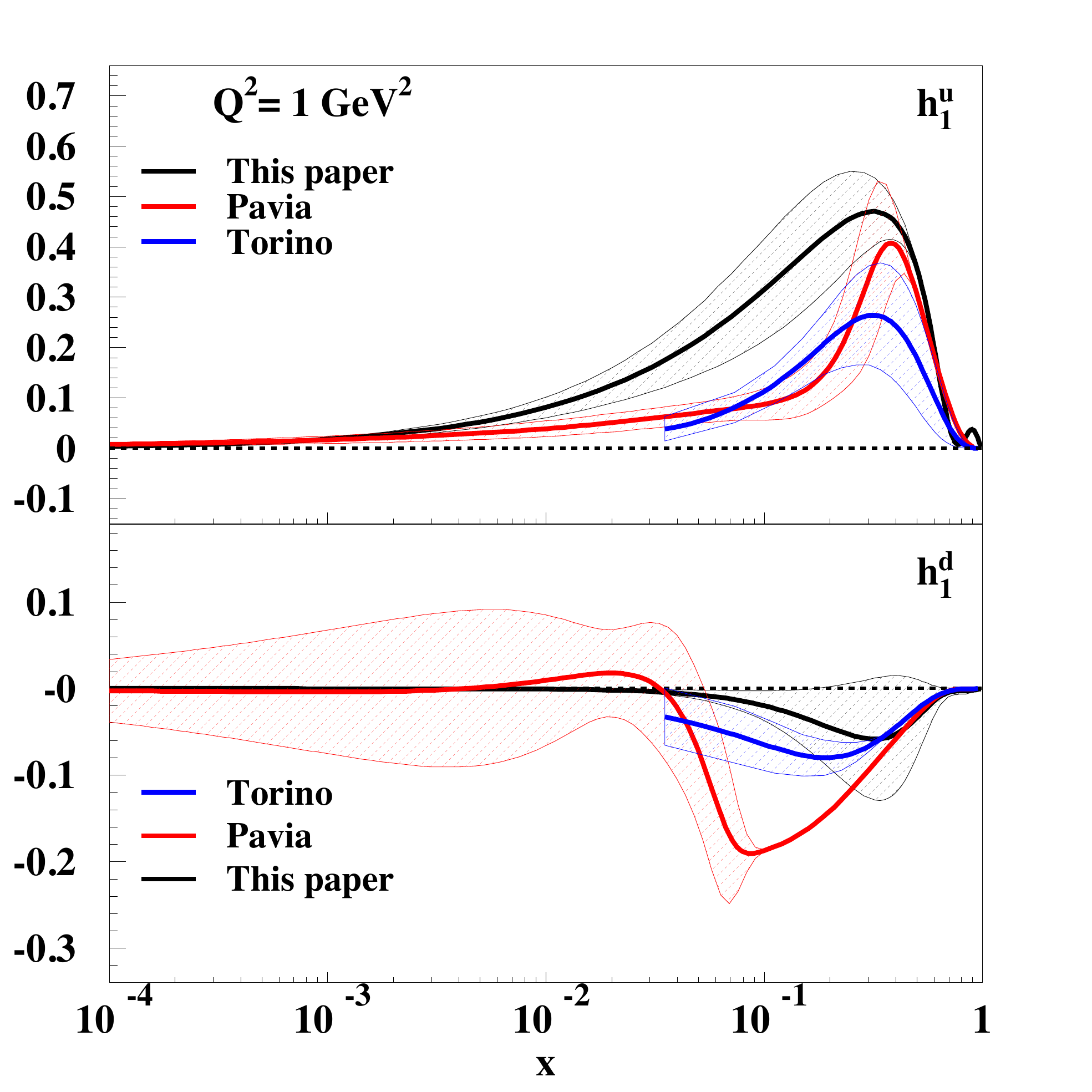}
\hspace{0.5cm}
\includegraphics[width=6.5cm]{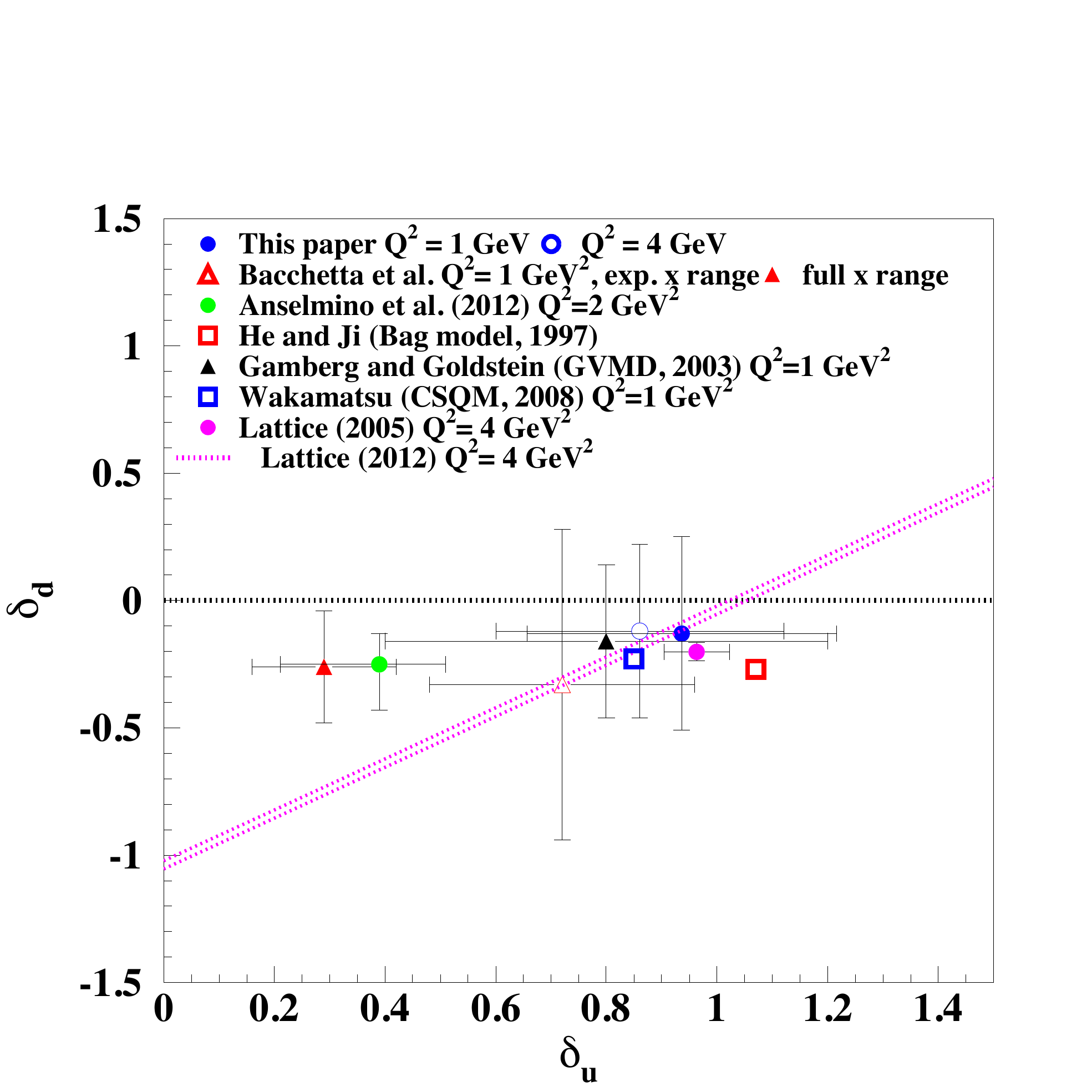}
\caption{Left (a): transversity, $h_1^q$ plotted vs. $x$ at $Q^2$= 1 GeV$^2$, for the $u$ quarks (upper panel) and for the $d$ quark (lower panel). Besides our analysis, the  recent extractions from  the Pavia group \cite{Courtoy} obtained from dihadron production in a collinear framework, and from the Torino group  \cite{Anselmino} obtained combining  data on polarized SIDIS single hadron production \cite{HERMES,COMPASS}, and  dihadron production from $e^+e^-$ annihilation \cite{Belle}. 
Right (b): Tensor charge values for the $d$ quark, $\delta_d$ plotted vs. the $u$ quark, $\delta_u$, as  obtained from our analysis of exclusive deeply virtual processes, from the other experimental extractions existing to date:  Pavia group \cite{Courtoy} ($Q^2$= 1 GeV$^2$, flexible set), and Torino group \cite{Anselmino}, and from different models. The thin band delimited by the dotted curves is  the lattice QCD result for the isovector component \cite{Engel} ($Q^2$= 4 GeV$^2$). For our model we also show the effect of PQCD evolution from $Q^2$=1 GeV$^2$ to $Q^2$= 4 GeV$^2$. Adapted from Ref.~\cite{GGL_eta}.}
\label{fig1}
\end{figure}

\begin{figure}
\includegraphics[width=6.5cm]{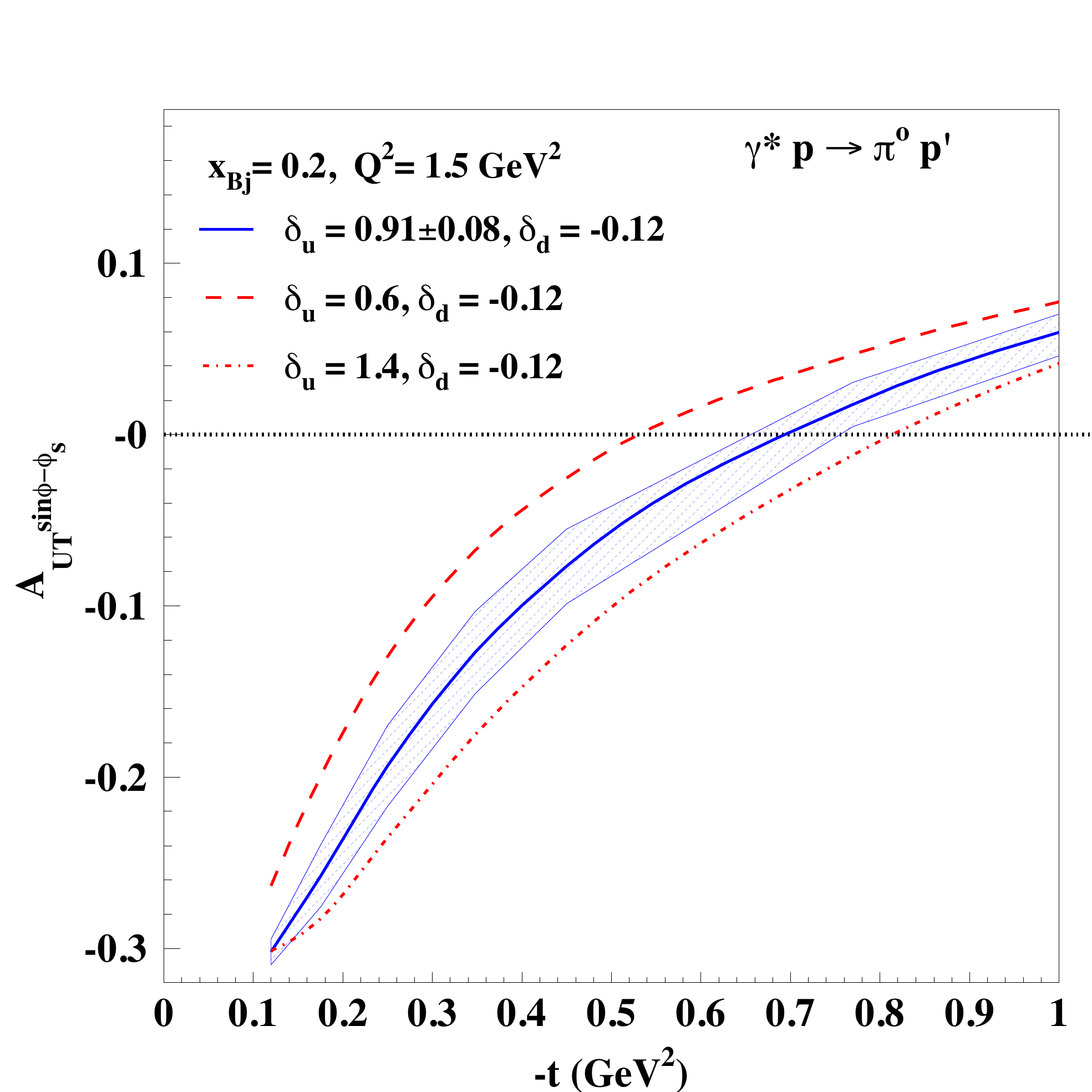}
\caption{(Color online) The asymmetry $A_{UT}^{\sin(\phi-\phi_S)}$, 
plotted vs. $-t$, at $x_{Bj}=0.2$, $Q^2=1.5$ for the $\gamma^* p \rightarrow \pi^0 p'$ reaction. The error band was obtained by varying the value of the $u$-quark tensor charge, $\delta_u$, by $\pm 0.08$. The dot-dashed curve corresponds to $\delta_u = 1.4$, and the dashed curve corresponds to $\delta_u=0.6$. The value of $\delta_d$ was kept fixed at $-0.12$. The graph shows the sensitivity of the asymmetry to variations of the tensor charge, or the precision that is needed in measurements of this quantity in order to reduce the size of the errors from the ones shown in Fig.
\protect\ref{fig1}. Adapted from Ref.~\cite{GGL_eta}}
\label{AUT_tensor}
\end{figure}

In advance of the development of the formalism and the brief presentation of our model, we show some important results - the transversity $h_1(x)$ in Fig~\ref{fig1}a, the tensor charges of the u and d quarks in Fig.~\ref{fig1}b, and the predicted asymmetry $A_{UT}^{\sin(\phi -\phi_s)}(x_{Bj},t, Q^2)$,  that is particularly sensitive to the transversity, in Fig.~\ref{AUT_tensor}.

DVCS  and DVMP  can be described within QCD factorization, through the convolution of specific GPDs and hard scattering amplitudes.  
There are four chiral-even GPDs, $H, E, \widetilde{H}, \widetilde{E}$ \cite{Ji_even} and
four additional chiral-odd GPDs, known to exist by considering twist-two quark operators that flip the net helicity by one unit, 
$H_T, E_T, \widetilde{H}_T, \widetilde{E}_T$ \cite{Ji_odd,Diehl_odd}.  
All GPDs depend on two additional kinematical invariants besides the parton's  Light Cone (LC) momentum fraction, $x$, and the DVCS process' four-momentum transfer, $Q^2$,
namely $t=\Delta^2$ where $\Delta=P-P'$ is the momentum transfer between the initial and final protons, and $\xi$, or the fraction of LC momentum transfer, $\xi=\Delta^+/(P^+ + P'^+)$. 
The observables containing the various GPDs are the so-called Compton Form Factors (CFFs), which are convolutions over $x$ of GPDs with the struck parton propagator. 

The quark GPDs are defined (at leading twist) as the matrix elements of  the following projection of the unintegrated quark-quark proton correlator (see Ref.\cite{Diehl_hab} for a detailed overview),
\begin{eqnarray}
W_{\Lambda', \Lambda}^\Gamma(x,\xi,t) & = & \frac{1}{2} \int \frac{d z^- }{2 \pi} e^{ix\overline{P}^+ z^-} \left. \langle P', \Lambda' \mid \overline{\psi}\left(-\frac{z}{2}\right) \Gamma \, \psi\left(\frac{z}{2}\right)\mid P, \Lambda \rangle \right|_{z^+=0,{\bf z}_T=0},
\label{matrix}
\end{eqnarray}
where $\Gamma=\gamma^+, \gamma^+\gamma_5, i\sigma^{i+}\gamma_5 (i=1,2)$, and the target's spins are $\Lambda, \Lambda^\prime$. 
For the two chiral-even cases
\begin{eqnarray}
W_{\Lambda', \Lambda}^{[\gamma^+]}(x,\xi,t)  & = &  \frac{1}{2\overline{P}^+} \overline{U}(P',\Lambda') \left( \gamma^{+} H(x,\xi,t) +
 \frac{i\sigma^{+\mu}(- \Delta_\mu)}{2M}  E(x,\xi,t)   \right)U(P,\Lambda); \label{correlator1} \\
W_{\Lambda', \Lambda}^{[\gamma^+\gamma^5]}(x,\xi,t)  & = &  \frac{1}{2\overline{P}^+} \overline{U}(P',\Lambda') \left( \gamma^{+} \gamma^5 {\tilde H}(x,\xi,t) +
\gamma^5 \frac{- \Delta^+}{2M}  {\tilde E}(x,\xi,t)   \right)U(P,\Lambda) 
\label{correlator2}
\end{eqnarray}


For the chiral-odd case, $\Gamma= i\sigma^{i+}\gamma_5$, $W_{\Lambda', \Lambda}^\Gamma$ was parametrized as \cite{Diehl_odd},
\begin{eqnarray}
\label{correlator}
W_{\Lambda', \Lambda}^{[i\sigma^{i+}\gamma_5]}(x,\xi,t)  & = &  \frac{1}{2\overline{P}^+} \overline{U}(P',\Lambda') \left( i \sigma^{+i} H_T(x,\xi,t) +
 \frac{\gamma^+ \Delta^i - \Delta^+ \gamma^i}{2M} E_T(x,\xi,t)   \right. \nonumber \\
& + & \left.  \frac{P^+ \Delta^i - \Delta^+ P^i}{M^2}  \widetilde{H}_T(x,\xi,t)  +
\frac{\gamma^+ P^i - P^+ \gamma^i}{2M} \widetilde{E}_T(x,\xi,t) \right) U(P,\Lambda)
\end{eqnarray}
The spin structures of GPDs that are directly related to spin dependent observables are most effectively expressed in term of helicity dependent amplitudes, developed extensively for the covariant description of two body scattering processes 
(see also Ref.\cite{Diehl_hab}).

So we see that there are 8 quark GPDs per flavor (at leading twist). Correspondingly, the quark-nucleon amplitudes have the form N$\rightarrow$ q :: q$^\prime \rightarrow$ N$^\prime$ with 8 independent helicity amps.
There are two questions to address:  How to model the 8 GPDs? How to measure them?
	     DVCS accesses Chiral Even GPDs through various cross sections and asymmetries. 
	     The GPDs, or their corresponding Compton Form Factors, enter linearly via Bethe-Heitler $\otimes$ DVCS interference.
DV$\pi^0$S accesses 2 Chiral Even + 4 Chiral Odd GPDs. This is a result of experimental observations that $d\sigma_T > d\sigma_L$ and the GPDs enter bilinearly via $d\sigma/d\Omega$ \& polarization asymmetries.

In  Ref.\cite{AGL},  after showing how  DV$\pi^0$P can be described in terms of chiral-odd GPDs, we estimated all of their contributions to the various observables with particular attention to the ones which were sensitive to the values of the tensor charge. Subsequently, 
an extension of the Flexible Reggeized Diquark model in the chiral-even sector in Refs.\cite{GGL,newFF} to the chiral odd sector was accomplished through the use of Parity and Charge Conjugation symmetry  relations obeyed by the various helicity structures in the model.
The chiral even GPDs are constrained to integrate to the nucleon  form factors,  
\begin{eqnarray}
\int_0^1 H^q(X,\zeta,t) & = F_1^q(t) \quad \int_0^1 E^q(X,\zeta,t) & = F_2^q(t) \\
\int_0^1 \widetilde{H}^q(X,\zeta,t) & = G_A^q(t) \quad \int_0^1 \widetilde{E}^q(X,\zeta,t) & = G_P^q(t) 
\label{GP}
\end{eqnarray}
where $F_1^q(t)$ and $F_2^q(t)$ are the Dirac and Pauli form factors for the quark $q$ components in the nucleon. $G_A^q(t)$ and $G_P^q(t)$ are the axial  and pseudoscalar form factors. 
Furthermore,  $H(x,0,0)=h_1(x)$ and $\widetilde{H}(x,0,0)=g_1(x)$. 

The connection of the correlator, Eq.(\ref{correlator1},\ref{correlator2},\ref{correlator}), with the helicity amplitudes proceeds by introducing \cite{AGL,GGL_pi0}, 
\begin{eqnarray}
f_{\Lambda_\gamma 0}^{\Lambda \Lambda^\prime} (\zeta,t)& = & \sum_{\lambda,\lambda^\prime} 
g_{\Lambda_\gamma 0}^{\lambda \lambda^\prime} (X,\zeta,t,Q^2)  \otimes
A_{\Lambda^\prime \lambda^\prime, \Lambda \lambda}(X,\zeta,t), 
\label{facto}
\end{eqnarray}
where the helicities of the virtual photon and the initial proton are, $\Lambda_\gamma$, $\Lambda$, 
and the helicities of  the produced pion and final proton are $0$, and $\Lambda^\prime$, respectively.
This describes a factorization into a ``hard part'', 
$g_{\Lambda_\gamma 0}^{\lambda  \lambda^\prime}$ for the partonic subprocess 
$\gamma^* + q \rightarrow \pi^0 + q$, 
and a ``soft part'' given by the quark-proton helicity amplitudes, $A_{\Lambda^\prime,\lambda^\prime;\Lambda,\lambda}$ 
that contain the GPDs.
The amplitudes $A_{\Lambda^\prime \lambda^\prime, \Lambda \lambda}$ implicitly contain an integration over the unobserved quark's transverse momentum, $k_T$,
and are functions of 
$x_{Bj} =Q^2/2M\nu \approx \zeta, t$ and $Q^2$. The convolution integral in  Eq.(\ref{facto}) 
 is given by $\otimes \rightarrow \int_{-\zeta+1}^1 d X$. 

The expressions for  the chiral-odd helicity amplitudes in terms of 
GPDs   \cite{Diehl_odd,Diehl_hab} are 
\begin{eqnarray}
\label{GPDodd}
A_{++,--} & = &  \sqrt{1-\xi^2}  \left[ { H}_ T + \frac{t_0-t}{4M^2} \widetilde{ H}_T      
- \frac{\xi^2}{1-\xi^2}  { E}_T  + \frac{\xi}{1-\xi^2} \widetilde{ E}_T \right]   \\ 
A_{+-,-+} & = &  - e^{-i2 \phi}  \sqrt{1-\xi^2}  \,  \frac{t_0-t}{4M^2} \, \widetilde{ H}_T \\
A_{++,+-} & = & e^{i \phi} \frac{\sqrt{t_0-t}}{4M} \left[ 2\widetilde{ H}_T  + (1-\xi)  \left({ E}_T + \widetilde{ E}_T \right) \right] \\ 
A_{-+,--} & = & e^{i \phi} \frac{\sqrt{t_0-t}}{4M}  \,  \left[  2\widetilde{ H}_ T + (1+\xi) \left( { E}_T - \widetilde{ E}_T \right) \right]  
\end{eqnarray}
where we use the symmetric  notation for the kinematic variables, $\phi$ is a phase given by the azimuthal angle of the vector {\bf D} with length $\mid {\bf D} \mid =\sqrt{t_o-t}/\sqrt{1-\xi^2}$. Analogous forms have been written for the chiral even sector \cite{Diehl_hab}.

For a transverse photon, inserting the expressions for $g_{10}^{+-}$ and the $A$'s into Eqs.(\ref{facto}) we obtain a set of helicity amplitudes that enter the observables. They include
\begin{eqnarray}
\label{helamps_gpd2}
f_{10}^{++} &= &  g_\pi^{V, odd}(Q)  e^{ i\phi}  \frac{\sqrt{t_0-t}}{4M} \left[ 2\widetilde{ \cal H}_T  + (1+\xi)  \left({\cal  E}_T - \widetilde{\cal  E}_T \right) \right]  \\
f_{10}^{+-} & = &   \frac{g_\pi^{V, odd}(Q)+ g_\pi^{A, odd}(Q)}{2} \, \sqrt{1-\xi^2}  \left[ { \cal H}_ T + \frac{t_0-t}{4M^2} \widetilde{ \cal H}_T      
- \frac{\xi^2}{1-\xi^2}  {\cal  E}_T  + \frac{\xi}{1-\xi^2} \widetilde{\cal  E}_T \right]   \\ 
 f_{10}^{-+} & = &  -   \frac{g_\pi^{A, odd}(Q)- g_\pi^{V, odd}(Q)}{2}  \, e^{- i 2 \phi}  \sqrt{1-\xi^2}  \,  \frac{t_0-t}{4M^2} \, \widetilde{\cal  H}_T  \\
f_{10}^{--} & = & g_\pi^{V, odd}(Q) e^{ i\phi} \frac{\sqrt{t_0-t}}{4M}  \,  \left[  2\widetilde{\cal H}_ T + (1-\xi) \left( { \cal E}_T + \widetilde{\cal E}_T \right) \right]  
\end{eqnarray} 
where ${\cal H}_T$, etc., are  the Compton form factors.
 In the $t$-channel picture, which has its roots in a Regge analysis of this process \cite{GolOwe}, one separates 
the $J^{PC}=1^{- -}$ and $J^{PC}=1^{+ -}$
contributions to the amplitudes for transverse and longitudinal virtual photons, respectively, thus generating two different types of $Q^2$ dependence at the $\pi^o$ 
vertex.

For a longitudinal photon one has the convolution of $g_{00}^{+-}$ with the $A$ helicity amplitudes . There are contributions for the longitudinal photon from the chiral even ${\cal H},\, {\cal E}$ as well (see Ref.~\cite{GGL_pi0}). Following from the chiral-even case, in $\pi^0$ electroproduction one obtains longitudinal photon amplitudes~\cite{VGG} that are significant. (see Ref.~\cite{GGL,GGL_pi0}).

Our model for evaluating the chiral-odd GPDs extends our reggeized diquark model, which was already configured for chiral-even GPDs, to the chiral-odd sector.  
We defined our approach as a  ``flexible parametrization" in that, mostly owing to its recursive feature, the different components can be efficiently fitted  separately as new data come in.  
The parameters were initially fixed by a fit applied recursively first to PDFs, and to the nucleon form factors. 
The model was shown to reproduce data on different observables in DVCS (charge \cite{HERMES1}, longitudinal  \cite{HallB} and transverse \cite{HERMES1} single spin asymmetries). A comparison with data from more recent analysis has also been shown in Ref.\cite{Seder}.  Recently \cite{newFF}, we presented a new fit that uses the form factor flavor separated data from Ref.\cite{Cates}. 

The basic structures  in our model are the quark-proton scattering amplitudes at leading order with proton-quark-diquark vertices.
The quark parton helicity amplitudes 
describe a two body process, $q^\prime(k^\prime) P \rightarrow q(k) P^\prime$,
where $q(k)$ corresponds to the ``struck quark". 
The intermediate diquark system, $X$, can have $J^P=0^+$ (scalar), or $J^P=1^+$ (axial vector). 
We start from the region $X\geq \zeta$. At fixed $M_X$, the amplitudes read for the Scalar diquark:
\begin{equation} 
\label{AmpS0}
A^{(0)}_{\Lambda^\prime \lambda^\prime, \Lambda \lambda}
  =  \int d^2k_\perp\phi^*_{\Lambda^\prime \lambda^\prime}(k^\prime,P^\prime) \phi_{\Lambda \lambda}(k,P).
\end{equation} 
We obtain for $S=0$ (the $S=1$ is given in Ref.~\cite{GGL_pi0}),  
\begin{eqnarray}
& \phi^*_{++}(k,P) & =  {\cal A}  \left(m+M X \right) =  \phi_{--}(k,P) ,           \\
& \phi^*_{+-}(k,P) & =  {\cal A}  ({k}_1 + i {k}_2) = -  \phi_{-+}(k,P),         
\end{eqnarray}

Next  we consider reggeization (see \cite{Forshaw,BroCloGun}, Ch.3 and references therein), that is, we extend the diquark model formalism to low $X$ by allowing the spectator system's mass 
to vary up to very large values.
This is accomplished by convoluting the GPD structures obtained in Eqs.(\ref{AmpS0}) with a ``spectral function", $\rho(M_X^2)$, where $M_X^2$ is the spectator's mass,  
\begin{eqnarray}
\label{reggeization}
F_T^q(X,\zeta,t) & = & {\cal N}_q  \int_0^\infty  d M_X^2  \rho(M_X^2)  F_{T}^{(m_q, M_\Lambda^q)}(X,\zeta,t; M_X). 
\end{eqnarray}
The spectral function was constructed in Refs.\cite{newFF,GGL} so that it approximately behaves as $(M_X^2)^{\alpha} \, {\rm for} \,\, M_X^2   \rightarrow \infty \,\, {\rm and} \,\,  \delta(M_X^2-\overline {M}_X^2) \, {\rm for} \,\,M_X^2$ at a  few $GeV^2$,
where $0 < \alpha <1$, and $\overline {M}_X$ is in the GeV range. Upon integration over the mass in Eq.(\ref{reggeization}) one obtains the desired $X^{-\alpha}$ behavior for small $X$, while for intermediate and large $X$ the integral is dominated by the $\delta$ function, yielding a result consistent with the diquark model (more details are given in Ref.\cite{newFF}). 

Inserting $\rho(M_X^2)$ in Eq.(\ref{reggeization}) one obtains an expression that we parametrized in a practical form as,
\begin{equation}
F_T^q(X,\zeta,t)   \approx    {\cal N}_q X^{-\alpha_q +\alpha'_q(X) t}   \,  F_{T}^{(m_q, M_\Lambda^q)}(X,\zeta,t; \overline{M}_X) = R^{\alpha_q,\alpha^\prime_q}_{p_q}(X,\zeta,t) \,  
 G_{M_X,m}^{M_\Lambda}(X,\zeta,t)
%
\label{fit_form}
\end{equation}
where $\alpha_q^\prime(X) =  \alpha_q^\prime  (1-X)^{p_q}$.
The  functions $G_{M_X,m_q}^{M_\Lambda^q}$ and $R^{\alpha_q,\alpha^\prime_q}_{p_q}$ are the quark-diquark and Regge contributions, respectively.

The fitting procedure of GPDs is quite complicated owing to its many different steps:
1) the construction of chiral-odd helicity amplitudes; 2) the connection of these amplitudes to the chiral-even ones using Parity relations within spectator models; 3) the fixing of chiral-even parameters at an initial scale, $Q_o^2$, using the nucleon form factors and PQCD evolution to match DIS data; 4) the determination of chiral-odd GPDs ; 5) the construction of the corresponding Compton form factors, and of the pseudoscalar meson electroproduction observables. 
In the following figures we show the GPDs at various kinematic ranges and some of the CFFs.



\begin{figure*}
\includegraphics[width=8.cm]{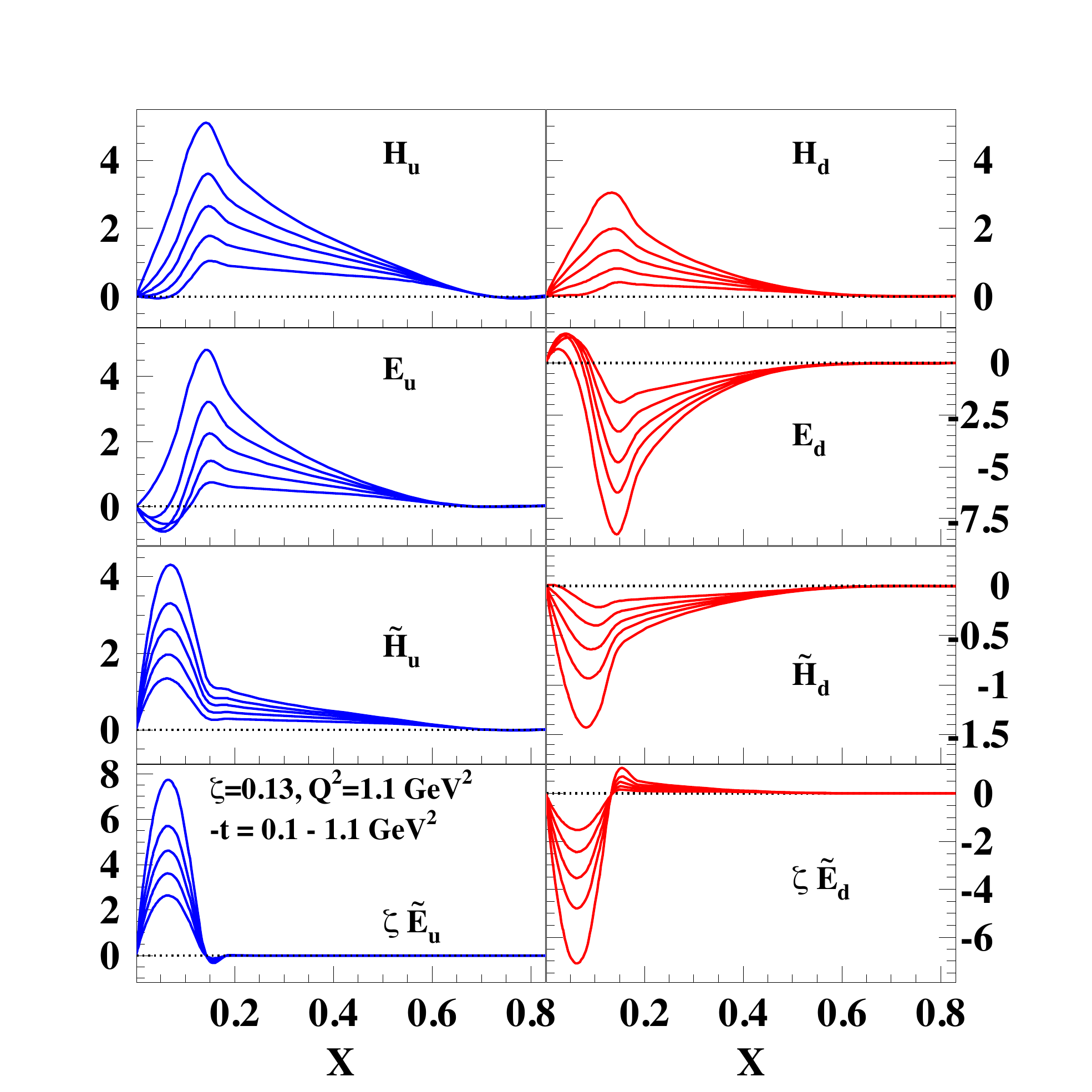}
\hspace{0.3cm}
\includegraphics[width=8.cm]{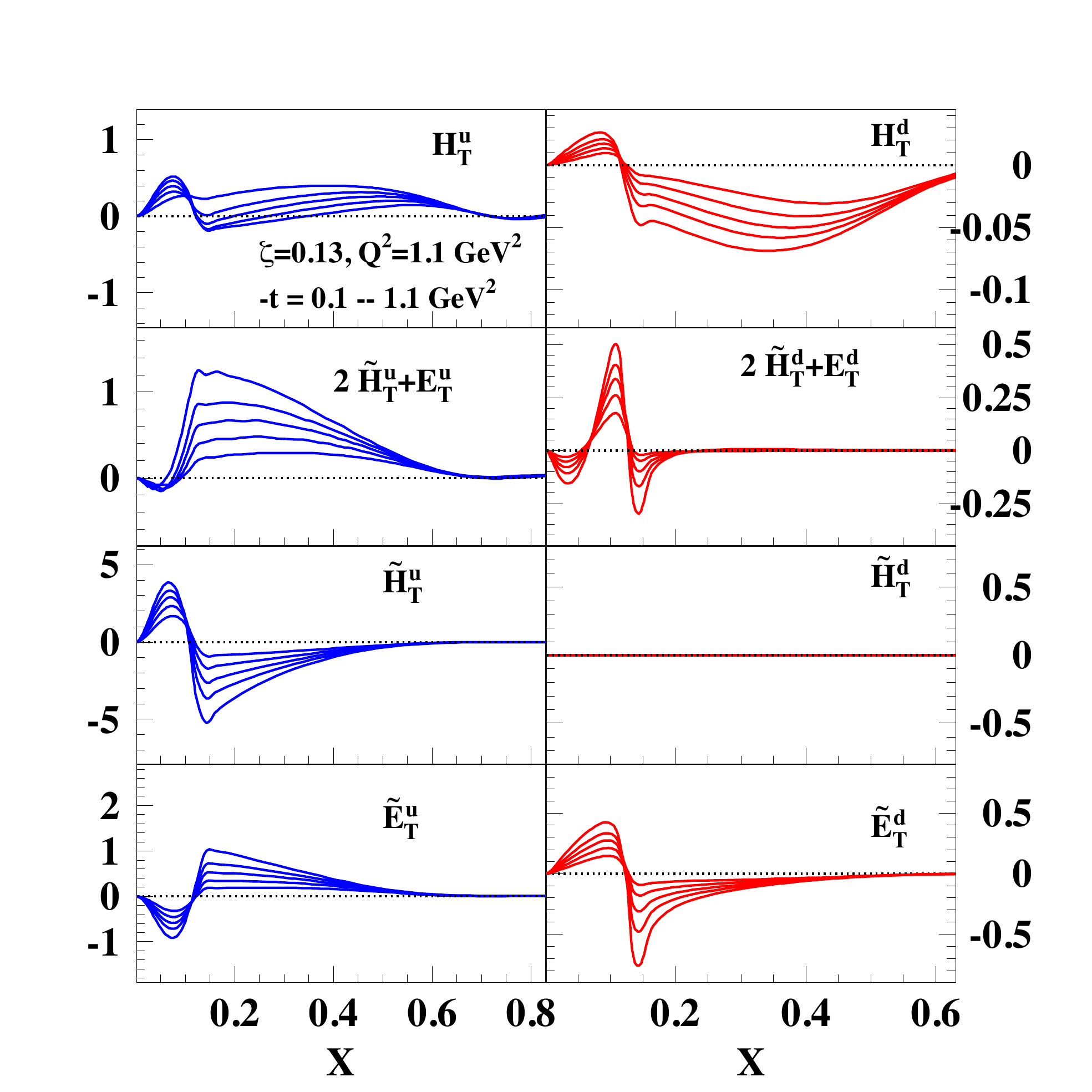}
\caption{(Color online) The chiral-even (left panel) and chiral-odd GPDs (right panel) evaluated using the model described in the text plotted vs. X at $x_{Bj}=\zeta=0.13$, $Q^2=  2$ GeV$^2$. The range in $-t$ is: $0.1  \leq -t \leq 1.1 \, {\rm GeV}^2$. Curves with the largest absolute values correspond to the lowest $t$.}
\label{fig:gpds_t}
\end{figure*}
A more detailed description of the other transversity functions including the first moment of $h_1^\perp \equiv 2\widetilde{H}_T^q + E_T^q$, whose integral over $X$ gives the transverse anomalous magnetic moments
\cite{Bur3}, is given in \cite{GGL_pi0}.

In Fig.\ref{fig:gpds_t} we show the $t$-dependent GPDs that enter the helicity amplitudes evaluated in Eqs.~\ref{GPDodd}
in a kinematical bin ($x_{Bj}=0.13, Q^2=1.1$ GeV$^2$) consistent with the Jefferson Lab kinematical coverage. 
The chiral-even GPDs are shown in the left panel, and the chiral-odd GPDs in the right panel.  

\begin{figure}
\includegraphics[width=9.cm]{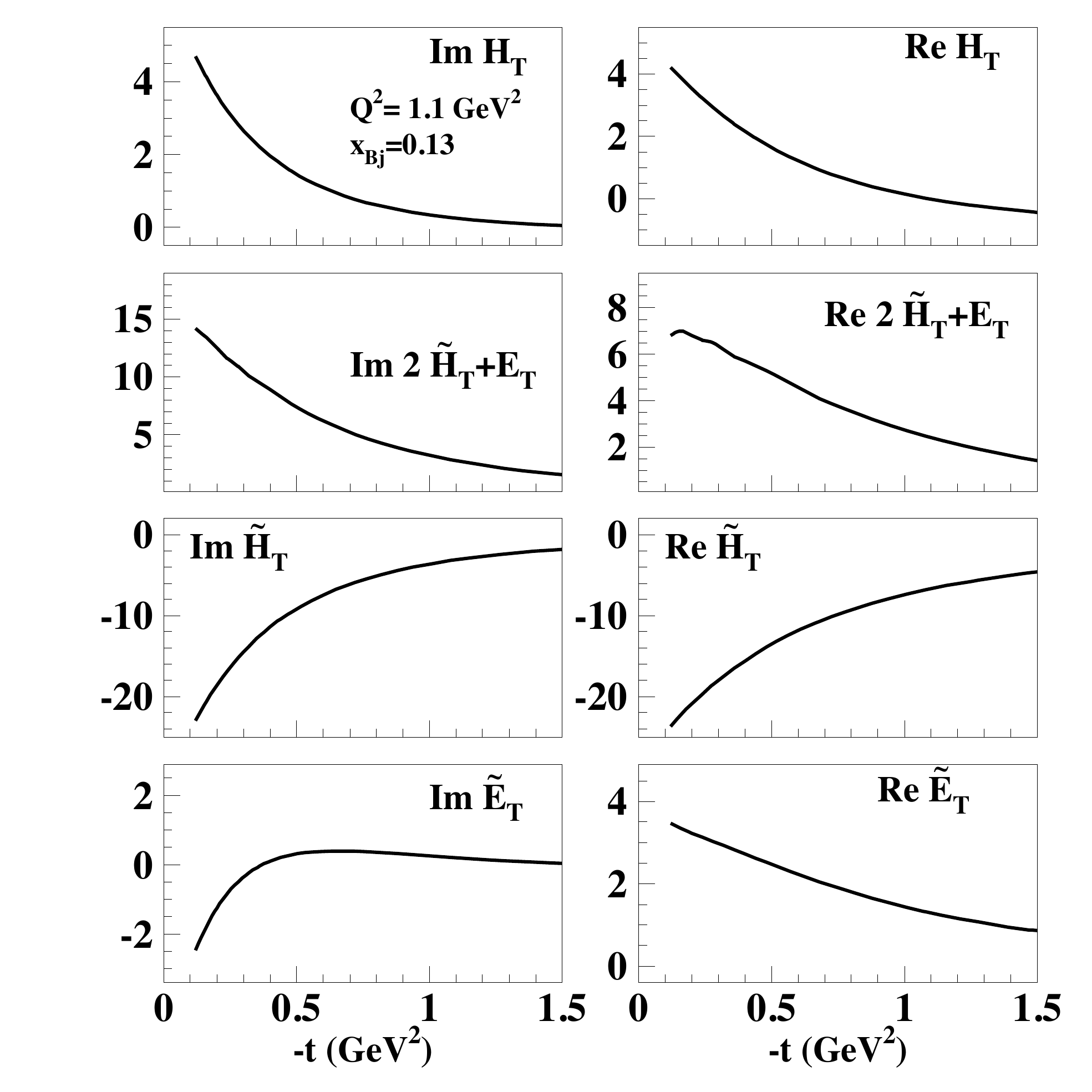}
\caption{Chiral-odd CFFs entering the process $\gamma^* p \rightarrow \pi^o p'$. From top to bottom  $\Im m{\cal H}_T$ (left), $\Re e{\cal H}_T$ (right);  $\Im m[2 \widetilde{\cal H}_T + {\cal E}_T]$ (left),  $\Re e[2 \widetilde{\cal H}_T + {\cal E}_T]$ (right); $\Im m \widetilde{\cal H}_T$ (left), $\Re e \widetilde{\cal H}_T$ (right); $\Im m \widetilde{\cal E}_T$ (left), $\Re e \widetilde{\cal E}_T$ (right). The various CFFs are  plotted vs. $-t$ for the kinematic bin $x_{Bj}=0.13$, $Q^2=1.1$ GeV$^2$.}
\label{fig:cff_odd}
\end{figure}

In Fig.\ref{fig:cff_odd} we show the proton CFFs, Eq.(\ref{helamps_gpd2}), which enter the $\gamma^* p \rightarrow \pi^o p'$ reaction. 
The various GPDs enter the helicity amplitudes and those, in turn, determine all the cross section terms for $\pi^0$ electroproduction.
\begin{figure}
\includegraphics[width=8.cm]{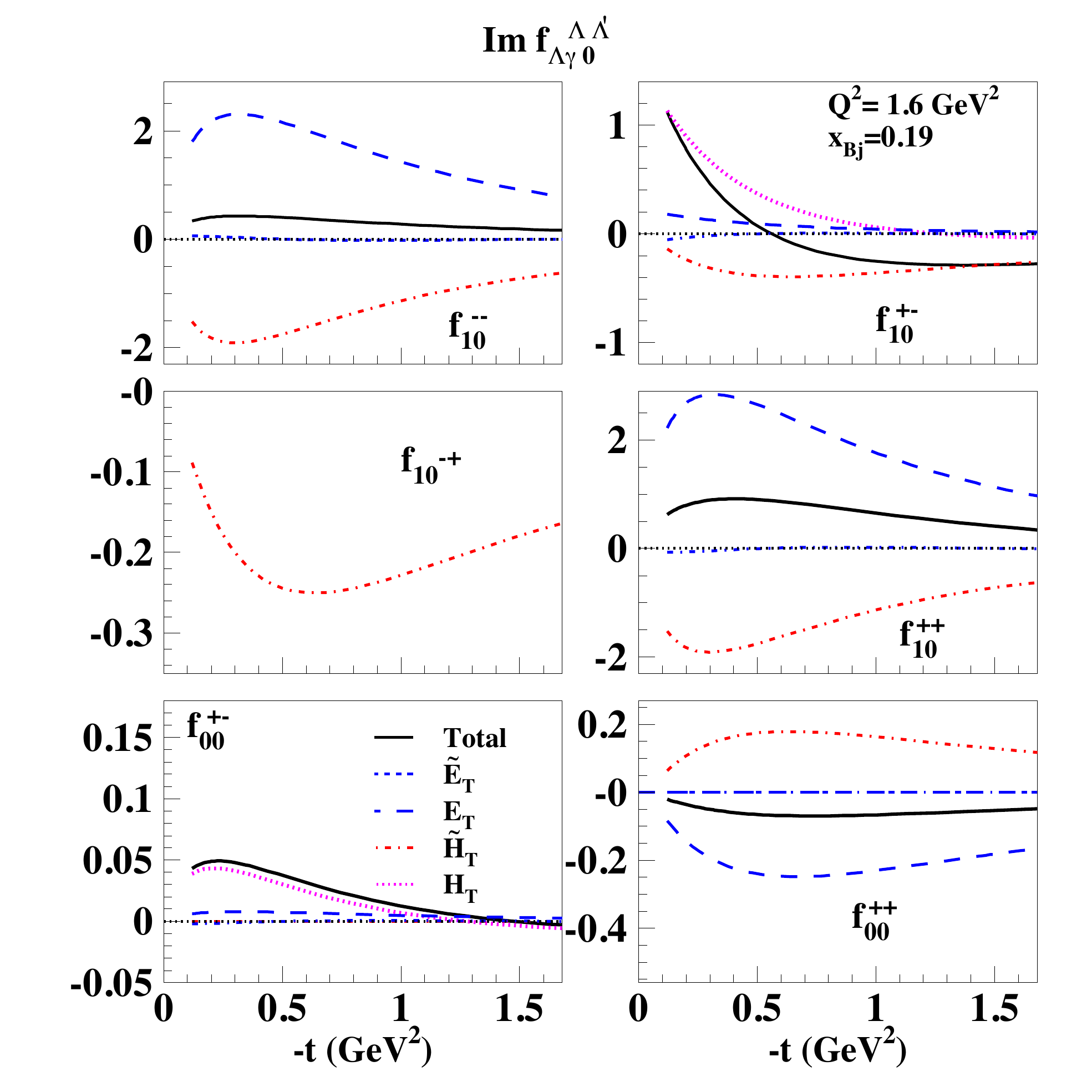}
\hspace{0.3cm}
\includegraphics[width=8.cm]{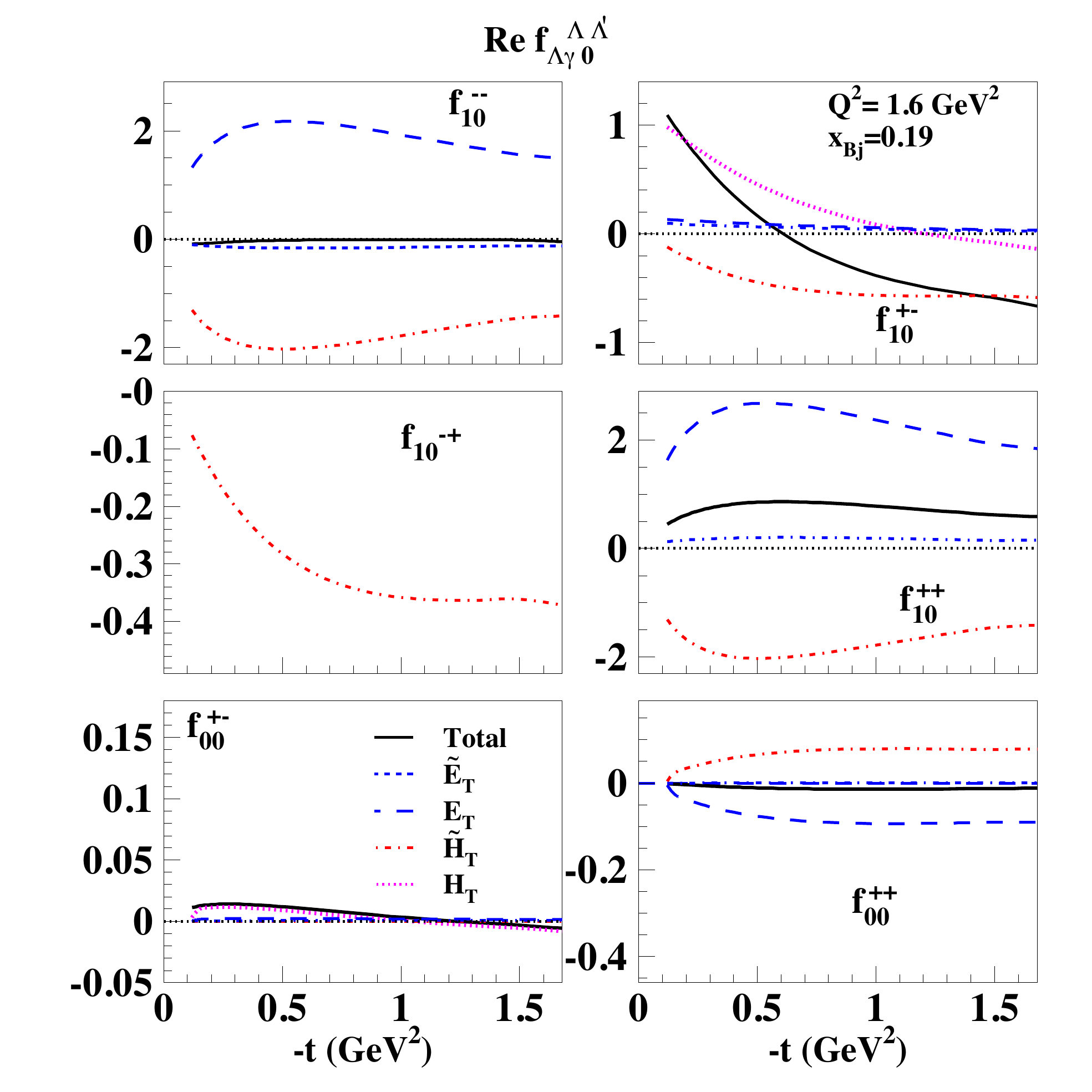}
\caption{(Color online) Helicity amplitudes for both transverse photon polarization, Eqs.(\ref{helamps_gpd2}), and longitudinal photon polarization, plotted vs. $-t$ for $x_{Bj} =0.19$, $Q^2= 1.6$ GeV$^2$. 
The imaginary parts are displayed on the left panel, and the real parts on the right panel.}
\label{fig:amps1}
\end{figure}

Helicity amplitudes are shown in Fig.\ref{fig:amps1} as functions of $-t$, for $x_{Bj}=0.19$, $Q^2 = 1.6$ GeV$^2$ (similar results are obtained for other kinematical bins in the range of Jefferson Lab data \cite{Kub}). 
The imaginary (real) parts are displayed on the LHS(RHS). The different contributions from the various chiral-odd GPDs,  are also shown in the figure.  

\noindent
(1) All GPDs contributions should be considered separately. In particular, $H_T$, $\widetilde{H}_T$, and $E_T$ are dominating; $\widetilde{E}_T$ is non zero in our model but small. 
Although the combination $2 \widetilde{H}_T + E_T$ might be considered more fundamental in that its spin structure corresponds to the Boer-Mulders function \cite{Diehl_hab}, and its first moment yields the proton's transverse anomalous magnetic moment \cite{Bur3}, $\widetilde{H}_T$, and $E_T$ appear separately, and multiplied by different factors in the amplitudes. $2 \widetilde{H}_T + E_T$ should just be viewed as a forward limit. 

\noindent
(2) The behavior of $f_{10}^{++}$ and $f_{10}^{--}$ is determined by $\widetilde{H}_T$, and $E_T$. As a consequence of  point (1), $f_{10}^{--}$ is sensibly different from $f_{00}^{++}$. In particular, because of the different multiplicative factors, $f_{10}^{--} <  f_{10}^{++}$.

\noindent 
(3) $f_{10}^{+-}$ is determined by $H_T$ at small $\mid t \mid$, and by $E_T$ at large $\mid t \mid$.

\noindent 
(4) $f_{10}^{-+}$ is determined by $\widetilde{H}_T$ only, but it is small due to the $\mid t \mid$ factor suppression. 

\noindent 
(5) The longitudinal photon contributions, $f_{00}^{+-}$, and $f_{00}^{++}$ are suppressed in the chiral-odd case.

\begin{figure}
\includegraphics[width=8.5cm]{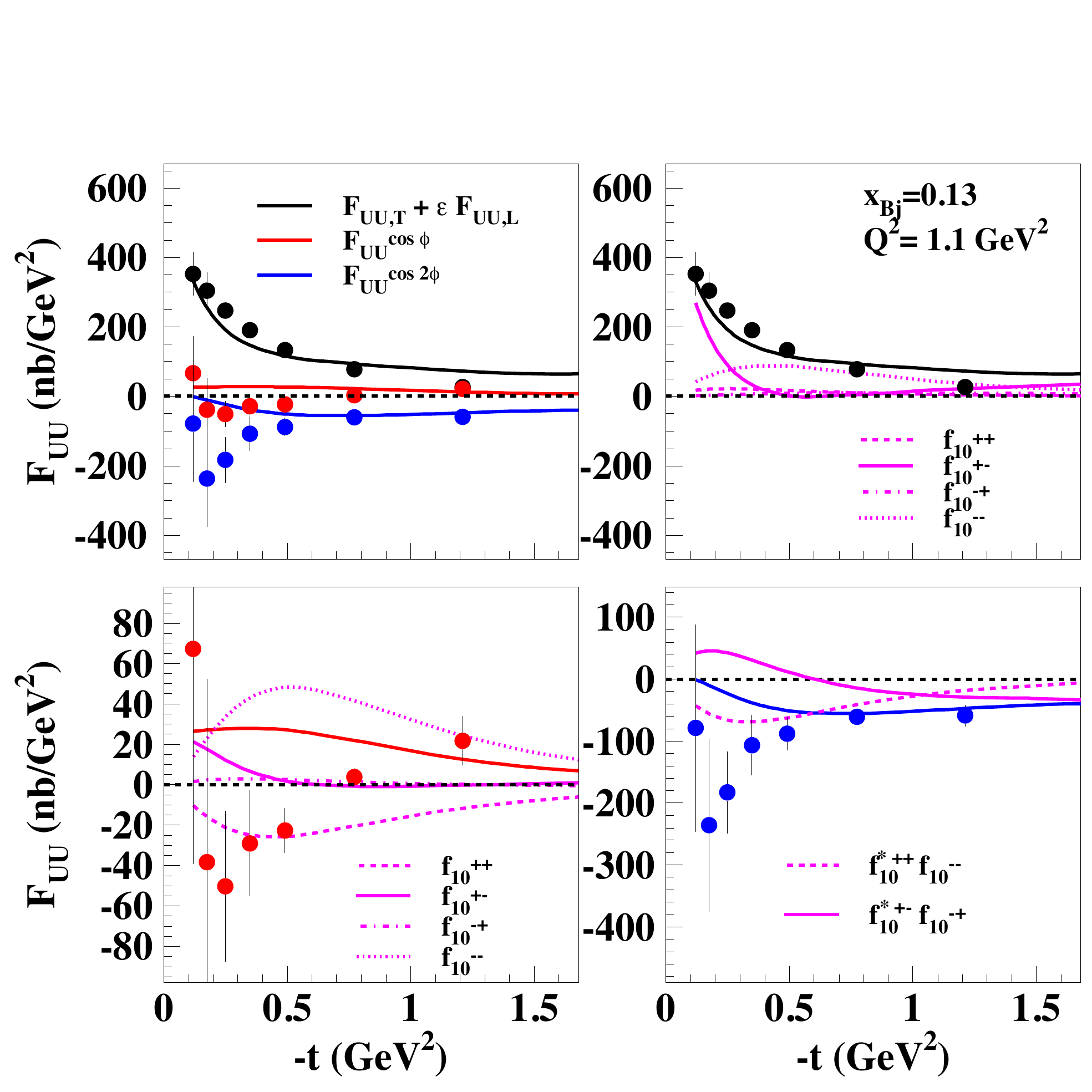}
\hspace{0.3cm}
\includegraphics[width=8.5cm]{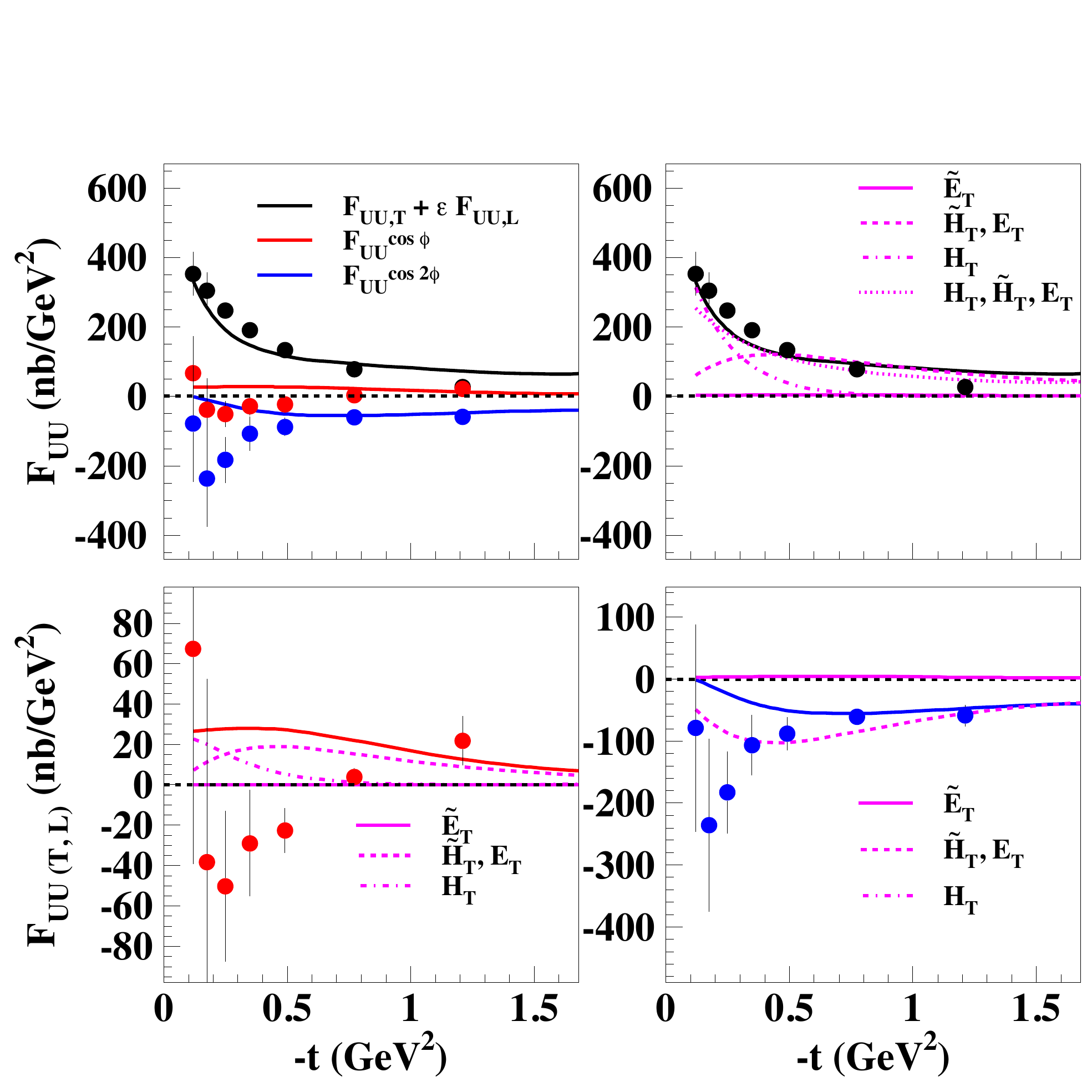}
\caption{(Color online) LEFT: Unpolarized cross section components, $F_{UU,T}+\epsilon \, F_{UU,L}$, $F_{UU}^{\cos 2 \phi}$, and $F_{UU}^{\cos \phi}$ in the kinematical bin, $x_{Bj} = 0.13$, $Q^2=1.2$ GeV$^2$. The upper left panel shows all components along with the data from Ref.\cite{Kub}.
The other panels show the contributions from the various helicity amplitudes. 
The right upper panel shows $F_{UU,T}+\epsilon \, F_{UU,L}$, and the contributions from $f_{10}^{++}$,  $f_{10}^{+-}$, $f_{10}^{-+}$  and $f_{10}^{--}$.  
Similarly, the lower left panel and the lower right panel  show the contributions of the various amplitudes to $F_{UU}^{\cos  \phi}$, and $F_{UU}^{\cos 2 \phi}$, respectively; 
RIGHT: Same as LEFT, displaying the GPDs components. The full curve is obtained by using only 
$\widetilde{E}_T$, the dashed curves by including only $2\widetilde{H}_T\pm (1\pm \xi)E_T$, the dot-dashed curve by including only $H_T$, and the dotted curve by including all GPDs, except for  $\widetilde{E}_T $. }
\label{fig_uu1}
\end{figure}

In Figure  \ref{fig_uu1}
we show unpolarized cross section components, $F_{UU,T}+\epsilon \, F_{UU,L}$, $F_{UU}^{\cos 2 \phi}$ , and $F_{UU}^{\cos \phi}$ as functions of $t$, for the kinematics $x_{Bj} = 0.13$, $Q^2=1.2$ GeV$^2$. The caption indicates the different contributions.
%

\begin{figure*}
\includegraphics[width=8.cm]{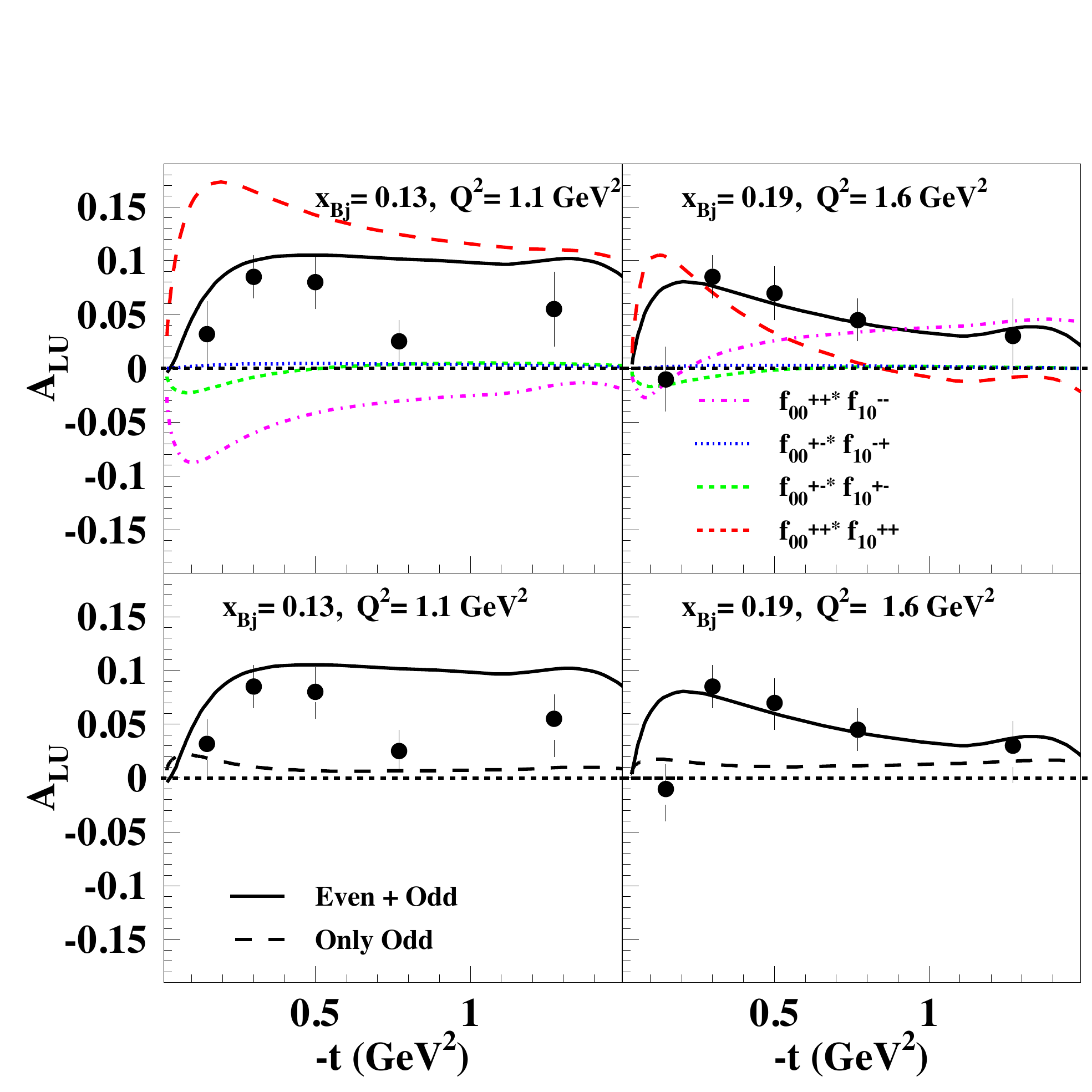}
\caption{(Color online) Beam spin asymmetry, $A_{LU}$, plotted vs. $-t$ for two different kinematics: $Q^2=1.1$ GeV$^2$,  
$x_{Bj}=0.13$ (left), $Q^2=1.6$ GeV$^2$,  
$x_{Bj}=0.19$ (right). 
Experimental data from Ref.\protect\cite{DeMasi:2008zz}. In the upper panels the different helicity amplitudes combinations contributing to $A_{LU}$ are shown.  The full curve describes the result obtained including all combinations.  In the lower panels we show results obtained including both the chiral-even and odd GPDs (full curve) compared to results obtained using the only the chiral-odd contribution (dashes). We conclude that the chiral-even GPDs dominate this observable.}
\label{fig:ALU}
\end{figure*}

The unpolarized $\sin \phi$ modulation, $F_{LU}^{\sin \phi}$ describes the beam asymmetry, $A_{LU}$,  
\begin{equation}
A_{LU}  = \sqrt{\epsilon(1-\epsilon)}  \, \frac{F_{LU}^{\sin \phi} }{F_{UU,T} + \epsilon F_{UU,L}}
\label{ALU}
\end{equation}
$A_{LU}$ is shown in Fig.\ref{fig:ALU} for two of the Jefferson Lab Hall B  kinematical bins along with the different amplitudes contributions, in this case the products: $(f_{10}^{++*} f_{00}^{++})$,  
$(f_{10}^{--*} f_{00}^{++})$, $(f_{10}^{-+*} f_{00}^{+-})$ and $(f_{10}^{+-*} f_{00}^{+-})$.
Notice that the longitudinally polarized amplitudes receive contributions from both the chiral-even and chiral-odd GPDs 
From the graph (lower panels) one can see a definite dominance of the chiral-even GPDs. We deduce that  $A_{LU}$ is not favored for the extraction of chiral-odd GPDs. The $A_{UT}$ in Fig.~\ref{AUT_tensor} is quite sensitive to the transversity, however.


Once established that the transversity parton distributions in the nucleon
can be accessed through deeply virtual exclusive pseudoscalar meson production which is sensitive to the chiral-odd transversity GPDs, $H_T, E_T, \widetilde{H}_T, \widetilde{E}_T$,
we addressed the feasibility of an experimental extraction.
This represents a consistent quantitative step with respect to our previous work \cite{AGL}.
In particular, only $H_T$ and the combination  $2\widetilde{H}_T+E_T$ \cite{Bur3}, which is related  to the first moment of the Boer-Mulders  TMD \cite{BoerMul}, were considered while $\widetilde{E}_T$, but now they are separated. A similar simplified approach was taken also in Ref.\cite{GolKro} - we differ in the importance attached to the skewedness dependence of $E_T, \, {\tilde E}_T$.

We see the results of our extended approach in relation to the many measured and measurable observables for deeply virtual pseudoscalar meson electroproduction. What is especially gratifying is that certain asymmetries constrain the GPDs well enough to separately determine $H_T$, and consequently  transversity through the limit $H_T(x,0,0)$, and the combination $2 \widetilde{H}_T + (1\pm \xi)  E_T$. 

\noindent {\bf Acknowledgments}
We thank the organizers of CIPANP2015 and Harut Avakian, Andrey Kim, Valery Kubarovsky and Paul Stoler for many useful discussions and suggestions.


\end{document}